# HSTS Preloading is Ineffective as a Long-Term, Wide-Scale MITM-Prevention Solution: Results from Analyzing the 2013 - 2017 HSTS Preload List

[1]JV Roig, [2]Eunice Grace V. Gatdula
[1] [2]Advanced Research Center – Asia Pacific College
[1]jvroig@gmail.com, [2]euniceg@apc.edu.ph

*Abstract*– HSTS (HTTP Strict Transport Security) serves to protect websites from certain attacks by allowing web servers to inform browsers that only secure HTTPS connections should be used. However, this still leaves the initial connection unsecured and vulnerable to man-in-the-middle attacks. The HSTS preload list, now supported by most major browsers, is an attempt to close this initial vulnerability. In this study, the researchers analyzed the HSTS preload list to see the status of its deployment and industry acceptance as of December 2017. The findings here show a bleak picture – adoption of the HSTS Preload List seem to be practically nil for essential industries like Finance, and a significant percentage of entries are test sites or nonfunctional.

## I. INTRODUCTION

HTTP Strict Transport Security (HSTS) is a web mechanism designed to protect websites from attacks such as man-in-the-middle attacks. This is accomplished by letting the HSTS-enabled server tell the user's browser that all connections to the site should be done only through secure, encrypted, HTTPS connections. This still leaves the initial connection unsecured and vulnerable to man-in-the-middle attacks[1], [2] – i.e, the very first time a browser connects to a website, before that specific website has the chance to inform the browser to only ever use secure HTTPS connections.

The HSTS preload list was created as an answer to this initial vulnerability window. This preload list is simply a list of websites that are preloaded into most major browsers. Whenever a user requests for a website that is in the preload list of the browser, the browser will automatically use only HTTPS connections, even before this site explicitly tells the browser to only use HTTPS. Effectively, this removes the vulnerability window in the initial request.

It has been over 5 years since the HSTS preload list has been first deployed in 2012. This study aims to analyze the current preload list as of the time of the study (Dec 2017) to see what important information can be gleaned as regards to its deployment and industry acceptance.

## II. BACKGROUND AND RELATED LITERATURE

The HSTS standard was published in 2012 by Jeff Hodges, Collin Jackson and Adam Barth through IETF RFC 6797 [3]. The original draft specification was in 2009, based on earlier work by Jackson and Barth in 2008 called "ForceHTTPS" [4].

Because the HSTS specification still leaves the initial connection unsecured, browser makers have adopted an additional feature as mitigation – the HSTS preload list [5][6][7]. The preload list is a list of sites that is built into major browsers. The browser always uses HTTPS to connect to these sites, as if those HSTS-enabled sites have already talked to the user's browser. This effectively "preloads" these sites, hence the name.

There is no significant published information regarding studies about the adoption rate or a breakdown of relevant statistics of the HSTS preload list. To our knowledge, this paper would be the first published article that presents an analysis of the growth, industry breakdown, and geographic breakdown of the HSTS preload list.

Troy Hunt, an Australian security professional and creator of the HaveIBeenPwned and Pwned Passwords services (https://haveibeenpwned.com/), published an article in his website in 2015 that lamented how few sites were registered in the preload list: "*… this is kinda depressing! Why? Because if we take Chromium's list as it stands today, there are less than three thousand sites worldwide demanding a secure connection and a significant portion of those are Google's.*" [8] Our interest in performing this study was inspired by this post from Hunt.

## III. RESEARCH DESIGN AND METHODOLOGY

### A. Overview

This research was done in several discrete steps:
1) All versions of the HSTS preload list until Dec 2017 was acquired through its public repository at the Chromium public repository [9]. The preload list is a file called *transport_security_state_static.json* inside the http folder of the repository. Cloning the repo locally gave the researchers access to the entire history of the preload list, every committed version from March 2013 to December 2017.
2) Historical HSTS preload list versions were used to trace the growth rate / acceptance rate of the HSTS preload list.
3) The latest version of the HSTS preload list was used for a more in-depth analysis – figuring out what sorts of industries were more represented in the list, as well as an attempt to profile geographic distribution.
4) The web presence of significant financial institutions in Asia was also matched against the preload list, to see how many among Asia's top financial institutions per country bothered to be included in the preload list.

All data was then summarized in tables and graphs, which are presented in section 3.

### B. Determining the Growth Rate

All versions of the HSTS preload list, starting from March 2013 to December 2017, were acquired through its public repository at the Chromium public repository [4]. For every succeeding month, starting from March 2013, the number of sites added to the HSTS preload list were counted to get an understanding of the eventual growth of the HSTS preload list, that is, from having 411 entries from March 2013 to having 41,308 entries by December 2017.

### C. Gathering Industry Data

The HSTS preload list at the end of December 2017 (containing 41,308 entries) was transformed into an Excel sheet. The information transferred to the Excel sheet contained basic preload list information:

- The list of individual HSTS preload entries as of December 2017
- The "force-https" setting
- Whether the subdomains of the sites are included in the preload list (as TRUE or FALSE)

Within the Excel sheet, the researchers created additional columns: *Country*, *Region*, and *Industry*. Country and Region indicate the geographic "home" of the website, according to language, domain, or explicit business/personal address or nationality that the owner expresses in the site. Industry indicates one of the fourteen categories we came up with to loosely classify sites being investigated:

1) Banking – sites that are affiliated with Banking organizations and may have online banking capabilities (such as transferring money and viewing your account balance.)
2) Finance – sites that are affiliated with organizations that deal with Insurance, Loans, Credit, Taxes, etc.
3) Retail – sites that specifically engage in the selling of products (such as clothing, appliances, make-up, medicine, etc.)
4) Company Website – sites that promote a company name or brand, exclusively. The official site of a company name or brand.
5) School/Education – sites that are affiliated with educational institutions, feature online educational courses, or deal with training, and the sharing of educational content.
6) Personal Website/ Blog – sites that promote a person or persons exclusively. The official site of a person, can include online portfolios or personal information and writing (such as blogs.)
7) Online Services – sites that provide a service online, that are not part of the School/Education category, such as photo-sharing sites, social media, gaming sites, video-sharing sites, emailing services, etc.
8) Streaming sites(illegal) – sites that provide streaming services of pirated video content from movies and TV shows.
9) Adult entertainment – sites involved in the sharing or streaming of pornographic content.
10) Not working – sites that could not be accessed (e.g., site can't be reached, bad gateway, Error 502 / 503 / 522, database errors, etc.)
11) Test website – sites that can be accessed but do not contain any substantial information, such as: Blank pages, sites that simply say 'Test', sites that only contain a gag image, etc.
12) NEWS – The official website of a NEWS organization, or websites that regularly publish objective write-ups written by various authors and contributors that tackle current affairs in Politics, Technology, Medicine, etc.
13) Government – The official website of a government organization. (e.g.,. Local police department, Department of Foreign Affairs, Local post offices, Office of the President, etc.)
14) Bitcoin – sites that promote and engage in the trade of Bitcoin or other online currencies.

A list of random numbers within the range of the HSTS December 2017 preload list (1 to 41,308) was utilized for the researchers to choose random entries from the preload list to investigate. The researchers visited each item or website from the HSTS list based on the given random numbers, and categorized each site's Country, Region, and Industry, whenever applicable. It took an estimated time of 100 hours to categorize 2,006 items in the HSTS preload list to their respective Country, Region and Industry (about 3 minutes on average for each site). This limit of 2,006 is simply due to the available time allotted for the study.

### D. Financial Institutions

The researchers focused on finding out if the official websites of known financial institutions throughout Asia, and their subdomains, are included in the December 2017 HSTS preload list. Specifically, the countries of Cambodia, China, East Timor, India, Indonesia, Iran, Hong Kong, Japan, Jordan, Laos, Lebanon, Macau, Malaysia, Mongolia, Nepal, North Korea, Oman, Pakistan, the Philippines, Qatar, Saudi Arabia, Singapore, South Korea, Sri Lanka, Syria, Taiwan, Tajikistan, Thailand, Turkey, Turkmenistan, United Arab Emirates, Uzbekistan, Vietnam, and Yemen were investigated.

The researchers gathered the names of 10 to 20 banks of each country from various financial websites and listings [10]-[43]. The researchers visited the official websites of the top 10 to 20 financial institutions of each country, and subsequently searched if these web addresses, and their subdomains, are present in the HSTS December 2017 preload list.

## IV. RESULTS AND ANALYSIS

### A. The HSTS Growth Rate

There was a gradual shift in the number of entries of the HSTS preload list from 2013 to 2017. As of 2015, the monthly number of entries of the HSTS preload list grew from being in the 10's to consistently being in the 100's. By 2016 to November 2017, the additional monthly number of entries of the HSTS preload list were consistently in the 1000's. Fig. 1 traces the annual growth rate of the HSTS preload list from 2013 to 2017. More detailed monthly views of the growth rate are available as supplementary information from https://research.jvroig.com/hsts_preload_2017

### B. Industry Data

After visiting 2,006 random HSTS preload list entries, and categorizing each site as one of the 14 given Industries for this study, Banking, Finance and Retail industries were found to be very underrepresented (industries which have the most to gain from HSTS preload listing). Regular company websites and personal blogs dominated the sample of entries analyzed. A significant portion of the sampled sites also seem to be Test sites (prototypes or work-in-progress or have been abandoned) or outright not working anymore. These findings are shown in Table 1.

### C. Country/Region Breakdown

The researchers also categorized the randomly sampled sites from the HSTS list according to country of origin or geographic region (with whatever precision the researchers could determine from examining the sites – country first, if available; then region if specific country information is not found; then "unknown" if no geographic info could be determined at all). Table 2 shows the results of this analysis. It is not surprising that the USA and major European countries (the "first-world"), make up most of the sampled entries in the HSTS preload list. However, over 700 sites (out of 2,006) could not be determined for country of origin or geographic region (labeled "Not Working" or "Unknown"), and the task of classifying random sites for actual into geographic locations is not very reliable. Consider this data set more as a curiosity, perhaps slightly informative, rather than incontrovertible evidence that participating in the HSTS preload list is a US- and Europe-centric activity.

### D. Financial Institutions

The top 10 to 20 financial institutions of 34 countries were reviewed to measure the acceptance of the HSTS preload list within the banking industry.

In total, 377 banks were identified, with 250 of them having dedicated websites. This list of 250 banking sites was then compared against the latest HSTS Preload List used in the study (Dec 2017). Dishearteningly, only 2 banks out of the 250 that the researchers tried to find were in the HSTS preload list, shown in Fig. 2.

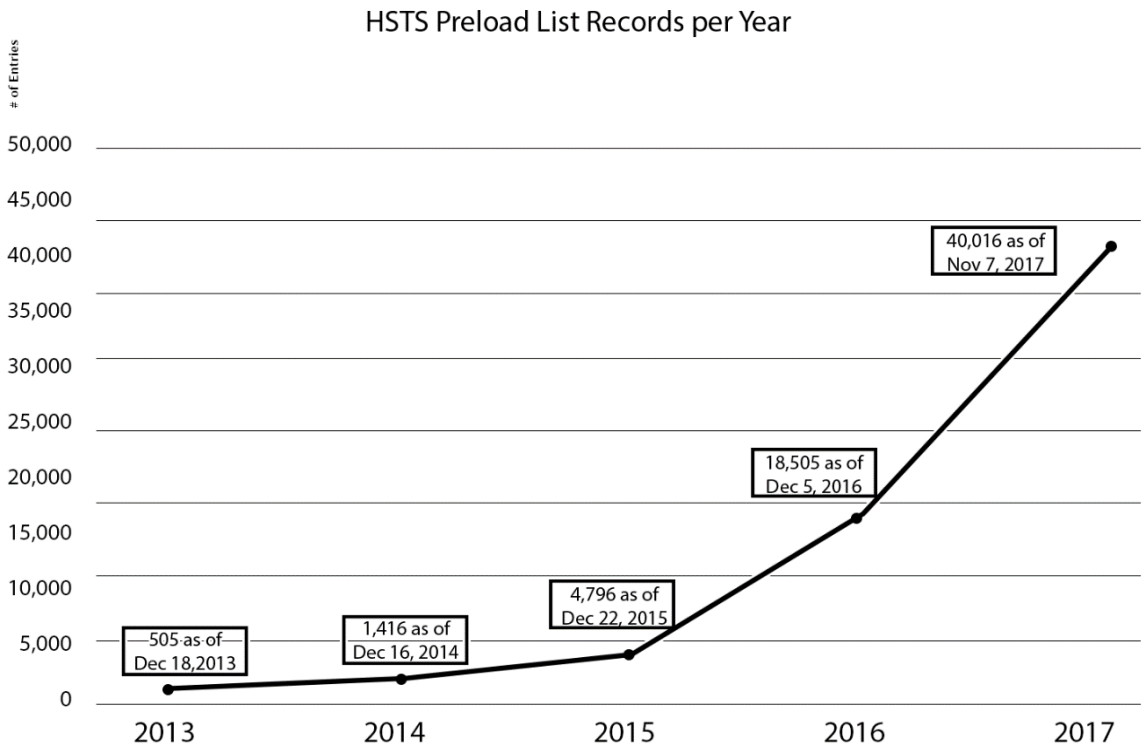

Fig. 1. HSTS Preload List Records Per Year

Table 1. HSTS Preload List Industry Data

| Industry | Count | Percentage |
|---|---|---|
| Banking | 5 | 0.25% |
| Finance | 38 | 1.89% |
| Retail | 98 | 4.89% |
| Company Website | 511 | 25.47% |
| School/Education | 30 | 1.50% |
| Personal Website/Blog | 448 | 22.33% |
| Online services | 220 | 10.97% |
| Streaming sites(illegal) | 5 | 0.25% |
| Adult entertainment | 3 | 0.15% |
| Not working | 403 | 20.09% |
| Test website | 206 | 10.27% |
| NEWS | 23 | 1.15% |
| Government | 12 | 0.60% |
| Bitcoin | 4 | 0.20% |
| **Total:** | **2,006** | **100.00%** |

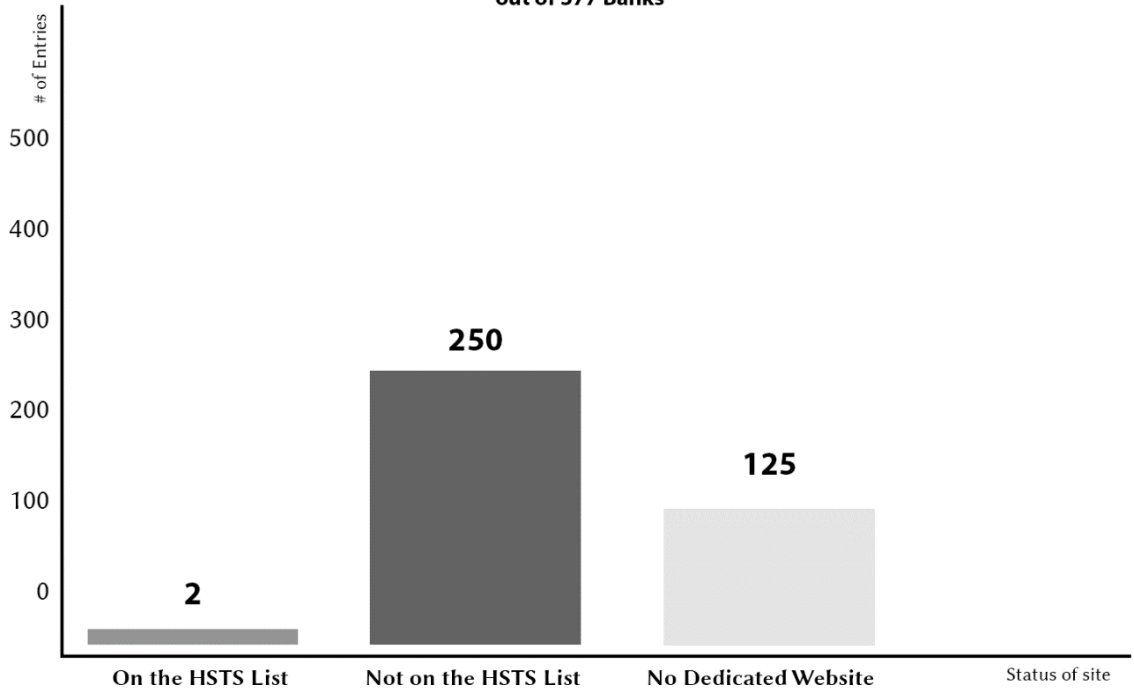

Fig. 2. HSTS Preload List investigation results of 377 Asian Banks

Table 2. HSTS Preload List Country/Region Data from random sample of 2,006 entries.

| Country | # | % | Country | # | % | Country | # | % |
|---|---|---|---|---|---|---|---|---|
| Africa | 1 | 0.05% | Greece | 5 | 0.25% | Poland | 11 | 0.55% |
| America (USA) | 226 | 11.27% | Holland | 1 | 0.05% | Portugal | 6 | 0.30% |
| Armenia | 1 | 0.05% | Hong Kong | 4 | 0.20% | Quebec | 1 | 0.05% |
| Australia | 34 | 1.69% | Hungary | 8 | 0.40% | Republic of Ireland | 6 | 0.30% |
| Austria | 14 | 0.70% | Iceland | 3 | 0.15% | Republic of Macedonia | 1 | 0.05% |
| Azerbaijan | 1 | 0.05% | India | 8 | 0.40% | Romania | 5 | 0.25% |
| Belgium | 10 | 0.50% | Indonesia | 12 | 0.60% | Rome | 4 | 0.20% |
| Bosnia | 1 | 0.05% | Israel | 1 | 0.05% | Russia | 37 | 1.84% |
| Brazil | 23 | 1.15% | Istanbul | 2 | 0.10% | Saudi Arabia | 1 | 0.05% |
| Bulgaria | 2 | 0.10% | Italy | 19 | 0.95% | Scotland | 4 | 0.20% |
| Canada | 20 | 1.00% | Japan | 59 | 2.94% | Serbia | 2 | 0.10% |
| Central America | 1 | 0.05% | Jordan | 1 | 0.05% | Slovakia | 6 | 0.30% |
| Chile | 1 | 0.05% | Korea | 2 | 0.10% | Slovenia | 1 | 0.05% |
| China | 44 | 2.19% | Lithuania | 3 | 0.15% | South Africa | 3 | 0.15% |
| Colombia | 1 | 0.05% | Luxembourg | 2 | 0.10% | Spain | 17 | 0.85% |
| Costa Rica | 2 | 0.10% | Malaysia | 3 | 0.15% | Sweden | 11 | 0.55% |
| Croatia | 2 | 0.10% | Mali | 1 | 0.05% | Switzerland | 43 | 2.14% |
| Cyprus | 3 | 0.15% | Malta | 1 | 0.05% | Taiwan | 3 | 0.15% |
| Czech Republic | 31 | 1.55% | Mexico | 5 | 0.25% | Thailand | 6 | 0.30% |
| Denmark | 24 | 1.20% | Netherlands | 73 | 3.64% | Tokelau | 3 | 0.15% |
| Deutschland | 4 | 0.20% | New Zealand | 10 | 0.50% | Tonga | 1 | 0.05% |
| England | 5 | 0.25% | Niue | 1 | 0.05% | Turkey | 8 | 0.40% |
| EU | 8 | 0.40% | Norway | 9 | 0.45% | UK | 149 | 7.43% |
| Europe | 2 | 0.10% | Not working | 403 | 20.09% | Ukraine | 5 | 0.25% |
| Finland | 6 | 0.30% | Palau | 2 | 0.10% | Unknown | 311 | 15.50% |
| France | 74 | 3.69% | Persia | 1 | 0.05% | Uruguay | 1 | 0.05% |
| Gabon | 1 | 0.05% | Peru | 1 | 0.05% | Venezuela | 2 | 0.10% |
| Germany | 169 | 8.42% | Philippines | 2 | 0.10% | Vietnam | 5 | 0.25% |

## V. CONCLUSION

The data found in this research paints a bleak picture for the HSTS preload list. Industries that are most at risk – banking and finance – have not been eager adopters of the HSTS preload list.

Why would this be the case? Not only is inclusion in the HSTS preload list absolutely free, it is just a few steps that would take no more than a couple of minutes. However, the researchers believe this is simply a natural and completely foreseeable effect of being opt-in: since inclusion in the HSTS preload list is "opt-in" (i.e., requires site admins to explicitly act to join) as opposed to "opt-out" (i.e., everyone is included by default, and requires site admins to explicitly act in order to not be included), the slow growth of the preload list can be seen as a natural consequence of the power of defaults.

It is also worth noting that with an estimate of over 1.7 billion websites in the world and over 200M unique domain names [44], having browsers maintain a preload list for HSTS is neither an ideal solution nor a scalable one. An ideal solution would be to simply have HTTPS as the default connection used by browsers and websites, foregoing plain HTTP unless a website specifically asked (in essence, a reverse HSTS setting, making it "opt-out"). This is not yet feasible worldwide, as many sites are still not HTTPS-ready, or are unable to function solely through HTTPS. Until then, the HSTS preload list is the best we've got against MITM attacks on HSTS-enabled servers. For now, while we wait for the time that the vast majority of the internet will happily work on pure, HTTPS-only protocol, more sites – especially those that handle confidential and financial records or transactions – should be exhorted to join the preload list.